\documentclass{jcgt}

\setciteauthor{TODO insert your name(s) here as they should appear in the citation}
\setcitetitle{TODO insert your paper's title here without linebreaks}
\accepted{submitted}{accepted}{published}{Editor Name}{vol}{issue}{1}{1}{year}
\seturl{http://jcgt.org/published/vol/issue/num/}

\usepackage{nccmath, amssymb}

\begin{document}

\title{Pulse Sequences to Observe NMR Coupled Relaxation in AX$_{\text{\bfseries\itshape n}}$ Spin Systems}

\author
       {Russell A. Brown}

\maketitle

\begin{abstract}
\small
NMR pulse sequences that are modifications of the HSQC experiment are proposed to observe ${}^{13}\textrm{C}$-coupled relaxation in AX, AX$_2$, and AX$_3$ spin systems. ${}^{13}\textrm{CH}$ and ${}^{13}{\textrm{CH}}_2$ moieties are discussed as exemplary AX and AX$_2$ spin systems. The pulse sequences may be used to produce 1D or 2D proton NMR spectra.
\end{abstract}

%-------------------------------------------------------------------------
\section*{Introduction}
\label{sec:introduction}
${}^{13}\textrm{C}$-coupled relaxation has been described in numerous publications. A representative publication provides an adequate overview of the discipline \cite{Werbelow}. Coupled-relaxation experiments permit estimation of the rate of molecular rotation in liquids via the Favro diffusion model \cite{Favro}. In addition, the rate of molecular conformational change may be estimated \cite{Ryabov}.

Estimates of the rates of molecular rotation and conformational change may be obtained via three steps. (1) Tentative estimates of the rates of rotation and conformational change are used to calculate spectral density functions \cite{Huntress}. (2) The spectral density functions are then used to calculate the elements of the Bloch-Redfield-Wangsness relaxation matrix \cite{Redfield}. (3) Nonlinear least squares are used to fit NMR spectra obtained via relaxation experiments to simulated spectra generated via solution of the Redfield differential equation. The least-squares fits refine the tentative estimates of the rotational diffusion coefficients and rates of conformational change.

${}^{13}\textrm{CH}$-coupled relaxation experiments have been used successfully to study the rotation of small molecules in liquids. However, attempts to extend these experiments to larger molecules such as polymers or peptides reveal that simulated spectra generated for an isolated ${}^{13}\textrm{CH}_2$ spin system that omits neighboring intramolecular protons lead to inaccurate least-squares fits \cite{Fuson}\cite{Brown1995}.

\newpage

Adding neighboring protons to the ${}^{13}\textrm{CH}_2$ spin system to extend it has been difficult because deriving equations that express the Redfield matrix elements in terms of spectral density functions for a given spin system is tedious \cite{Zheng} and because the Redfield matrix grows as $4^{s}$, where $s$ represents the number of spins in the extended system, which requires inordinate computation to solve the Redfield differential equation via matrix diagonalization for any values of $s$ except small values.

These two limitations have been overcome by a novel reformulation of the relaxation theory equations that avoids computationally expensive matrix diagonalization and hence permits extension of the spin system by adding neighboring protons \cite{Kuprov2011}\cite{Kuprov2021}. Moreover, the embodiment of this reformulation in the \emph{Spinach} library \cite{Spinach} obviates the need for tedious derivation of equations that express the Redfield matrix elements in terms of spectral density functions. These innovations motivate the proposal of five pulse sequences that permit observation of ${}^{13}\textrm{C}$-coupled relaxation via proton NMR spectra of AX, AX$_2$, and AX$_3$ spin systems. The pulse sequences are presented below for ${}^{13}\textrm{CH}$ and ${}^{13}{\textrm{CH}}_2$ moieties.

\section*{Carbon-Proton Multiple Quantum Relaxation Pulse Sequence}
\label{sec:CarbonProtonMultipleQuantum}

\begin{figure}[h]
\centering
\centerline{\includegraphics[width=5.15in]{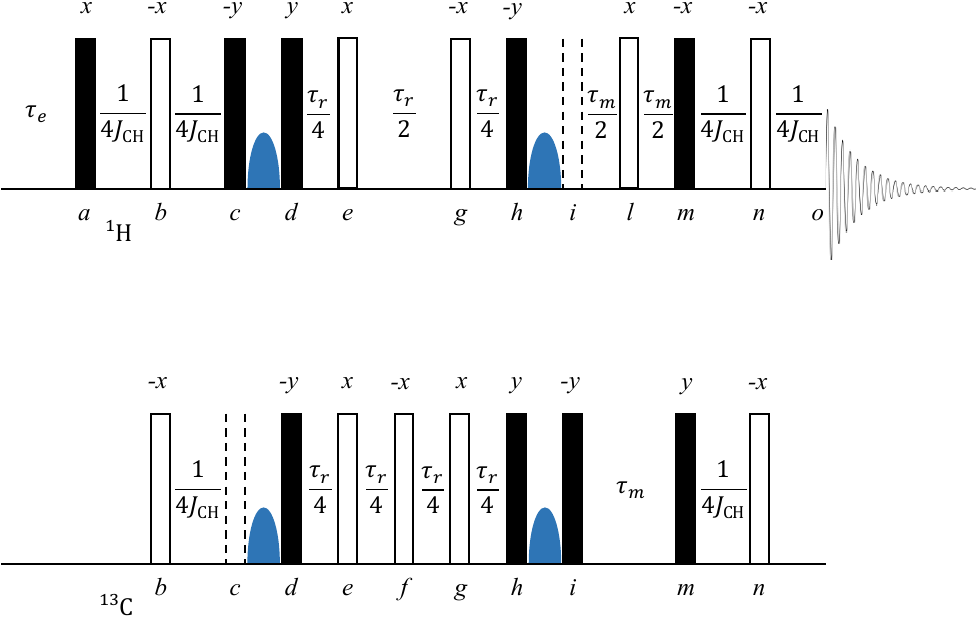}}
\caption{Carbon-Proton Multiple Quantum Relaxation Pulse Sequence}
\label{fig:CarbonProtonMultipleQuantumPulseSequence}
\end{figure}

The pulse sequence depicted in Figure \ref{fig:CarbonProtonMultipleQuantumPulseSequence} observes relaxation of carbon-proton multiple quantum coherences for ${}^{13}\textrm{CH}$. It inserts features of an HMQC experiment \cite{Bax} into an HSQC experiment \cite{Bodenhausen}.

\newpage

 In Figure \ref{fig:CarbonProtonMultipleQuantumPulseSequence}, $90^{\circ}$ and $180^{\circ}$ pulses are depicted by black and white rectangles respectively, the carbon-proton scalar coupling constant is designated by $J_{\mathrm{CH}}$, and a purge gradient is depicted by a blue sync function \cite{Keeler1}.
 
 The pulse sequence begins with a delay time $\tau_e$ that allows the spin system to achieve the thermal equilibrium state $4\hat{I_z}+\hat{S_z}$ where $I$ and $S$ represent the proton and carbon spins respectively. After the $\tau_e$ delay, a proton $90_{x}^{\circ}$ pulse creates $4\hat{I_y}$ in-phase proton single quantum coherences (SQCs) that dephase to $2\hat{I_x}\hat{S_z}$ anti-phase SQCs during time interval $a$-$c$. A proton $90_{-y}^{\circ}$ pulse at time $c$ creates a $-2\hat{I_z}\hat{S_z}$ $J$-ordered state (aka a $zz$ state) \cite{Morris}\cite{Burum}.

A purge gradient applied during time interval $c$-$d$ suppresses the $4\hat{I_y}$ SQCs for all protons that are not coupled to ${}^{13}\textrm{C}$ \cite{Keeler2}. These protons experience a spin echo during time interval $a$-$c$ that refocuses $4\hat{I_y}$ instead of dephasing it to $2\hat{I_x}\hat{S_z}$. The $4\hat{I_y}$ SQCs are unaffected by the proton $90_{-y}^{\circ}$ pulse at time $c$, so they persist and are suppressed by the purge gradient.

At time $d$, proton $90_{y}^{\circ}$ and carbon $90_{-y}^{\circ}$ pulses convert $-2\hat{I_z}\hat{S_z}$ to $2\hat{I_x}\hat{S_x}$ carbon-proton multiple quantum coherences (MQCs). These pulses also convert the carbon $\hat{S_z}$ to $-\hat{S_x}$ SQCs that produce no signals observable in the proton spectrum at time $o$. Then the spin system relaxes for a variable delay time $\tau_{r}$ during time interval $d$-$h$.

Proton and carbon $180_{\pm{x}}^{\circ}$ pulses at times $e$, $f$, and $g$ refocus at time $h$ the effects of the proton and carbon chemical shifts and the $J_{\mathrm{CH}}$ scalar coupling. The proton and carbon $180_{\pm{x}}^{\circ}$ pulses at times $e$ and $g$ are optional for the AX but required for the AX$_2$ and AX$_3$ spin systems wherein the MQCs evolve under the influence of $J_{\mathrm{CH}}$.

Assuming no relaxation at time $h$, the $2\hat{I_x}\hat{S_x}$ MQCs persist. Assuming full relaxation at time $h$, the $z$-magnetization for any proton is $4\hat{I_z}$, independent of whether that proton is coupled to ${}^{13}\textrm{C}$. For intermediate relaxation between the extremes of no relaxation and full relaxation, the magnetization is a combination of $2\hat{I_x}\hat{S_x}$ and $4\hat{I_z}$.

At time $h$, proton $90_{-y}^{\circ}$ and carbon $90_{y}^{\circ}$ pulses reconvert $2\hat{I_x}\hat{S_x}$ to  $-2\hat{I_z}\hat{S_z}$ and convert $4\hat{I_z}$, which developed due to relaxation, to $4\hat{I_x}$ SQCs that are suppressed by a purge gradient applied during time interval $h$-$i$.

At time $i$, a carbon $90_{-y}^{\circ}$ pulse converts $-2\hat{I_z}\hat{S_z}$ to $2\hat{I_z}\hat{S_x}$. Then the spin system evolves for a variable delay time $\tau_{m}$ during time interval $i$-$m$. A proton $180_{x}^{\circ}$ pulse at time $l$ refocuses the effects of the proton chemical shift and the $J_{\mathrm{CH}}$ scalar coupling so that $2\hat{I_z}\hat{S_x}$ evolves by $\exp \left ( -i \omega_{\mathrm{C}} \tau_{m} \right )$ to become the following at time $m$
\begin{equation}
2\exp \left ( -i \omega_{\mathrm{C}} \tau_{m} \right ) \hat{I_z}\hat{S_x} = 2\cos \left ( \omega_{\mathrm{C}} \tau_{m} \right ) \hat{I_z}\hat{S_x} - 2\sin \left ( \omega_{\mathrm{C}} \tau_{m} \right ) \hat{I_z}\hat{S_y}
\label{eq:Exp}
\end{equation}
 where $\omega_{\mathrm{C}}$ is the carbon Larmor frequency.
 
 At time $m$, proton $90_{-x}^{\circ}$ and carbon $90_{y}^{\circ}$ pulses convert $2\cos \left ( \omega_{\mathrm{C}} \tau_{m} \right ) \hat{I_z}\hat{S_x}$ to $-2\cos \left ( \omega_{\mathrm{C}} \tau_{m} \right )\hat{I_y}\hat{S_z}$ anti-phase proton SQCs. (A carbon $90_{x}^{\circ}$ pulse instead of a carbon $90_{y}^{\circ}$ pulse creates $-2\sin \left ( \omega_{\mathrm{C}} \tau_{m} \right )\hat{I_y}\hat{S_z}$ SCQs.) During time interval $m$-$o$, a rephasing pulse sequence converts  $-2\cos \left ( \omega_{\mathrm{C}} \tau_{m} \right )\hat{I_y}\hat{S_z}$ to $4\cos \left ( \omega_{\mathrm{C}} \tau_{m} \right )\hat{I_x}$.
 
For relaxation delays $\tau_r=0$ and $\tau_r=\infty$, the $\alpha$ and $\beta$ components of the proton transverse magnetization on the $x$-axis (i.e., the proton doublet) at time $o$ are

\begin{equation}
\begin{matrix}
 \alpha_{\tau_r={ \scriptscriptstyle 0} } = 2\cos \left ( \omega_{\mathrm{C}} \tau_{m} \right ) & \alpha_{\tau_r=\infty}= 0
 \\ 
 \beta_{\tau_r={ \scriptscriptstyle 0} } = 2\cos \left ( \omega_{\mathrm{C}} \tau_{m} \right ) & \beta_{\tau_r=\infty}= 0
\end{matrix}
\label{eq:Cos1}
\end{equation}

It is possible to modify the pulse sequence of time interval $m$-$o$ to create a sensitivity-enhanced 2D experiment \cite{Keeler3} and to modify the pulse sequence of time interval $i$-$o$ to create a 2D TROSY experiment \cite{Pervushin}\cite{Keeler4}.

\section*{Proton Longitudinal Relaxation Pulse Sequence}
\label{sec:ProtonLongitudinal}

\begin{figure}[h]
\centering
\centerline{\includegraphics[width=5.15in]{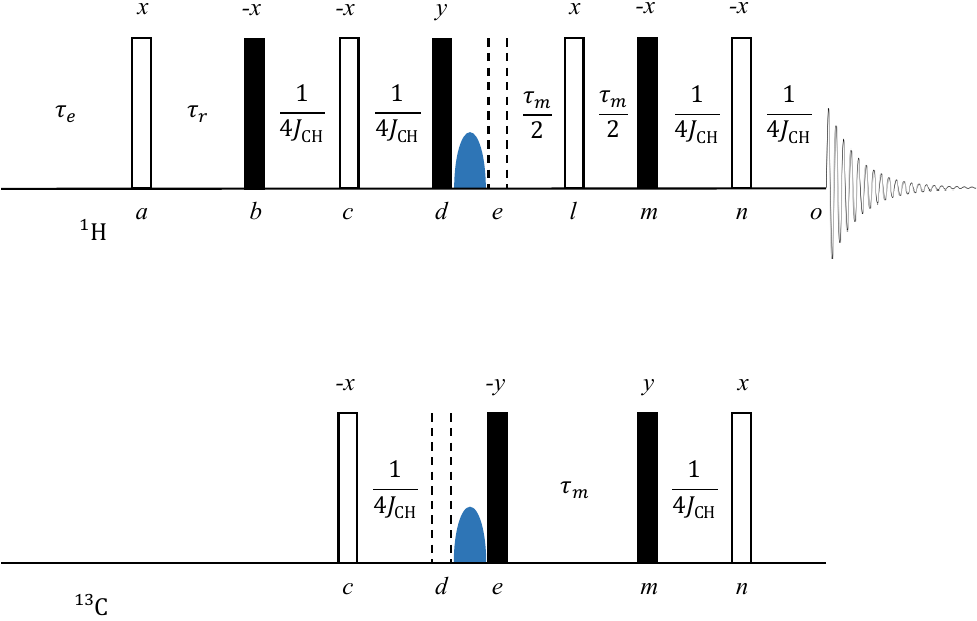}}
\caption{Proton Longitudinal Relaxation Pulse Sequence}
\label{fig:ProtonLongitudinalPulseSequence}
\end{figure}

The pulse sequence depicted in Figure \ref{fig:ProtonLongitudinalPulseSequence} observes proton longitudinal relaxation for ${}^{13}\textrm{CH}$. It performs proton inversion recovery followed by an HSQC experiment. It begins with a delay time $\tau_e$ that allows the spin system to achieve the thermal equilibrium state $4\hat{I_z}+\hat{S_z}$. At time $a$, a proton $180_{x}^{\circ}$ pulse inverts $4\hat{I_z}$ to obtain $-4\hat{I_z}$. Then the spin system relaxes for a variable delay time $\tau_{r}$ during time interval $a$-$b$.

A dephasing pulse sequence applied during time interval $b$-$e$ creates $\pm 2\hat{I_z}\hat{S_x}$ anti-phase carbon SQCs at time $e$ and suppresses $4\hat{I_y}$ SQCs for any proton that is not coupled to ${}^{13}\textrm{C}$. The remainder of the proton longitudinal relaxation pulse sequence applied during time interval $e$-$o$ achieves a similar result to time interval $i$-$o$ of the carbon-proton multiple quantum relaxation pulse sequence depicted in Figure \ref{fig:CarbonProtonMultipleQuantumPulseSequence}.

For relaxation delays $\tau_r=0$ and $\tau_r=\infty$, the $\alpha$ and $\beta$ components of the proton transverse magnetization on the $x$-axis (i.e., the proton doublet) at time $o$ are

\begin{equation}
\begin{matrix}
 \alpha_{\tau_r={ \scriptscriptstyle 0} } = -2\cos \left ( \omega_{\mathrm{C}} \tau_{m} \right ) & \alpha_{\tau_r=\infty}= 2\cos \left ( \omega_{\mathrm{C}} \tau_{m} \right )
 \\ 
 \beta_{\tau_r={ \scriptscriptstyle 0} } = -2\cos \left ( \omega_{\mathrm{C}} \tau_{m} \right ) & \beta_{\tau_r=\infty}= 2\cos \left ( \omega_{\mathrm{C}} \tau_{m} \right )
\end{matrix}
\label{eq:Cos2}
\end{equation}

\section*{Proton Transverse Relaxation Pulse Sequence}
\label{sec:ProtonTransverse}

\begin{figure}[h]
\centering
\centerline{\includegraphics[width=5.15in]{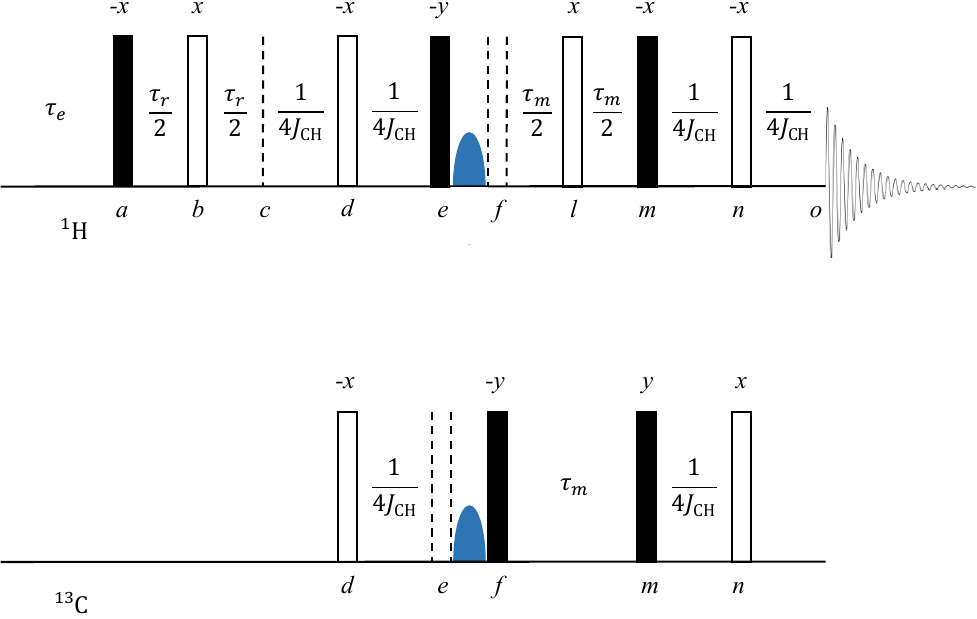}}
\caption{Proton Transverse Relaxation Pulse Sequence}
\label{fig:ProtonTransversePulseSequence}
\end{figure}

The pulse sequence depicted in Figure \ref{fig:ProtonTransversePulseSequence} observes proton transverse relaxation for ${}^{13}\textrm{CH}$. It performs a proton spin echo followed by an HSQC experiment. It begins with a delay time $\tau_e$ that allows the spin system to achieve the thermal equilibrium state $4\hat{I_z}+\hat{S_z}$. At time $a$, a proton $90_{-x}^{\circ}$ pulse converts $4\hat{I_z}$ to $4\hat{I_y}$ SQCs. Then the spin system relaxes for a variable delay time $\tau_{r}$ during spin--echo time interval $a$-$c$.

A dephasing pulse sequence applied during time interval $c$-$f$ creates $-2\hat{I_z}\hat{S_x}$ anti-phase carbon SQCs at time $f$ and suppresses $4\hat{I_y}$ SQCs for any proton that is not coupled to ${}^{13}\textrm{C}$. The remainder of the proton transverse relaxation pulse sequence applied during time interval $f$-$o$ achieves a similar result to time interval $i$-$o$ of the carbon-proton multiple quantum relaxation pulse sequence depicted in Figure \ref{fig:CarbonProtonMultipleQuantumPulseSequence}.

For relaxation delays $\tau_r=0$ and $\tau_r=\infty$, the $\alpha$ and $\beta$ components of the proton transverse magnetization on the $x$-axis (i.e., the proton doublet) at time $o$ are

\begin{equation}
\begin{matrix}
 \alpha_{\tau_r={ \scriptscriptstyle 0} } = 2\cos \left ( \omega_{\mathrm{C}} \tau_{m} \right ) & \alpha_{\tau_r=\infty}= 0
 \\ 
 \beta_{\tau_r={ \scriptscriptstyle 0} } = 2\cos \left ( \omega_{\mathrm{C}} \tau_{m} \right ) & \beta_{\tau_r=\infty}= 0
\end{matrix}
\label{eq:Cos3}
\end{equation}

\section*{Carbon Transverse Relaxation Pulse Sequence}
\label{sec:CarbonTransverse}

\begin{figure}[ht]
\centering
\centerline{\includegraphics[width=5.15in]{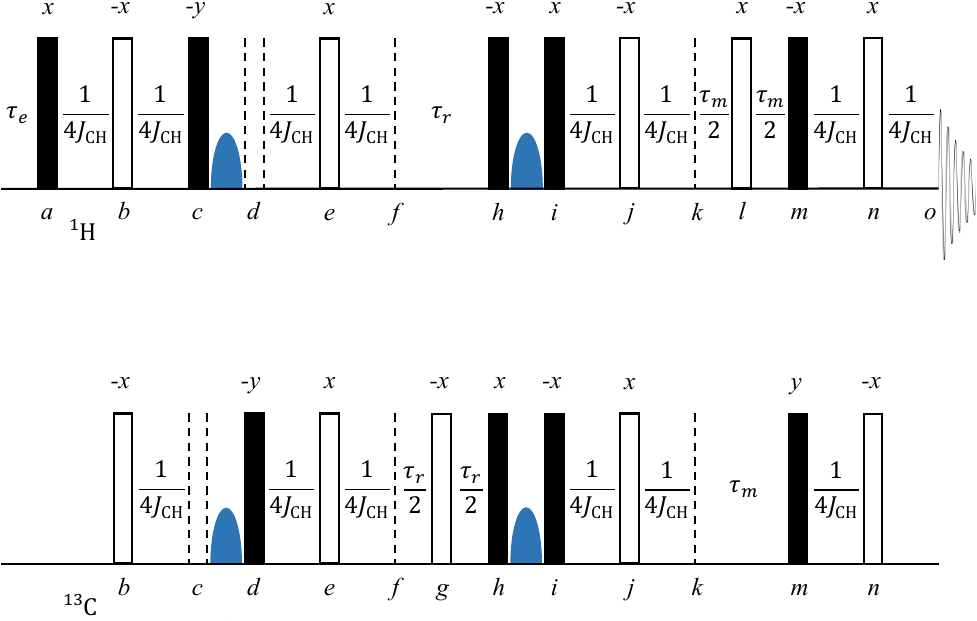}}
\caption{Carbon Transverse Relaxation Pulse Sequence}
\label{fig:CarbonTransversePulseSequence}
\end{figure}

The pulse sequence depicted in Figure \ref{fig:CarbonTransversePulseSequence} observes carbon transverse relaxation for ${}^{13}\textrm{CH}$. It inserts a carbon spin echo into an HSQC experiment. It begins with a delay time $\tau_e$ that allows the spin system to achieve the thermal equilibrium state $4\hat{I_z}+\hat{S_z}$.

During time interval $a$-$d$, a dephasing pulse sequence creates $-2\hat{I_z}\hat{S_x}$ and $-4\hat{S_x}$ and suppresses $4\hat{I_y}$ for any proton not coupled to ${}^{13}\textrm{C}$. During time interval $d$-$f$, a rephasing pulse sequence converts $-2\hat{I_z}\hat{S_x}$ and $-4\hat{S_x}$ to $4\hat{S_y}$ and $-\hat{I_z}\hat{S_y}$. During spin-echo interval $f$-$h$, the spin system relaxes for a variable delay time $\tau_{r}$. At time $h$, proton $90_{-x}^{\circ}$ and carbon  $90_{x}^{\circ}$ pulses convert $4\hat{S_y}$ and $-\hat{I_z}\hat{S_y}$ to $4\hat{S_z}$ and $-\hat{I_y}\hat{S_z}$ and convert $4\hat{I_z}$, which developed due to relaxation, to $4\hat{I_y}$. During time interval $h$-$i$, a purge gradient suppresses all SQCs. At time $i$, a carbon $90_{-x}^{\circ}$ pulse reconverts $4\hat{S_z}$ to $4\hat{S_y}$. (A proton $90_{x}^{\circ}$ pulse at time $i$ is optional for AX but required for AX$_2$ and AX$_3$ to reconvert to $\hat{S_y}$ the proton-proton MQCs created by the proton $90_{-x}^{\circ}$ pulse at time $h$.) During time interval $i$-$k$, a dephasing pulse sequence reconverts $4\hat{S_y}$ to $-4\hat{I_z}\hat{S_x}$. The remainder of the carbon transverse relaxation pulse sequence applied during time interval $k$-$o$ achieves a similar result to time interval $i$-$o$ of the carbon-proton multiple quantum relaxation pulse sequence depicted in Figure \ref{fig:CarbonProtonMultipleQuantumPulseSequence}.

For relaxation delays $\tau_r=0$ and $\tau_r=\infty$, the $\alpha$ and $\beta$ components of the proton transverse magnetization on the $x$-axis (i.e., the proton doublet) at time $o$ are 

\begin{equation}
\begin{matrix}
 \alpha_{\tau_r={ \scriptscriptstyle 0} } = 2\cos \left ( \omega_{\mathrm{C}} \tau_{m} \right ) & \alpha_{\tau_r=\infty}= 0
 \\ 
 \beta_{\tau_r={ \scriptscriptstyle 0} } = 2\cos \left ( \omega_{\mathrm{C}} \tau_{m} \right ) & \beta_{\tau_r=\infty}= 0
\end{matrix}
\label{eq:Cos4}
\end{equation}

The rephasing and dephasing pulse sequences applied during time intervals $d$-$f$ and $i$-$k$ preserve at time $k$ the $-2\hat{I_z}\hat{S_x}$ carbon SQCs created at time $d$. If these pulse sequences were omitted, then $-2\hat{I_z}\hat{S_x}$ would persist at time $h$ and be converted to $-2\hat{I_y}\hat{S_x}$ MQCs by the proton $90_{-x}^{\circ}$ pulse at time $h$. Then the purge gradient of time interval $h$-$i$ would suppress the double-quantum-coherence component of those MQCs. In that case, the carbon $90_{-x}^{\circ}$ pulse at time $i$ would not restore $-2\hat{I_z}\hat{S_x}$.

For the AX$_2$ spin system, the optimum delays during time intervals $d$-$f$ and $i$-$k$ are $ \frac{1} {8J_\mathrm{CH}} $ instead of $ \frac {1} {4J_\mathrm{CH}} $. For $ \frac{1} {8J_\mathrm{CH}} $ delays, the $\alpha$ and $\beta$ components of the proton transverse magnetization at time $o$ for relaxation delays $\tau_r=0$ and $\tau_r=\infty$ are

\begin{equation}
\begin{matrix}
 \alpha_{\tau_r={ \scriptscriptstyle 0} } = \frac {3} {2} \cos \left ( \omega_{\mathrm{C}} \tau_{m} \right ) & \alpha_{\tau_r=\infty}= 0
 \\ 
 \beta_{\tau_r={ \scriptscriptstyle 0} } = \frac {3} {2} \cos \left ( \omega_{\mathrm{C}} \tau_{m} \right ) & \beta_{\tau_r=\infty}= 0
\end{matrix}
\label{eq:Cos5}
\end{equation}
Because the purge gradient of time interval $h$-$i$ suppresses the proton-proton double quantum coherences created at time $h$, $ \alpha_{\tau_r={ \scriptscriptstyle 0} } = \beta_{\tau_r={ \scriptscriptstyle 0} } < 2\cos \left ( \omega_{\mathrm{C}} \tau_{m} \right )  $ at time $o$.

\section*{Proton-Proton Multiple Quantum Relaxation Pulse Sequence}
\label{secProtonProtonMultipleQuantum}

\begin{figure}[h]
\centering
\centerline{\includegraphics[width=5.55in]{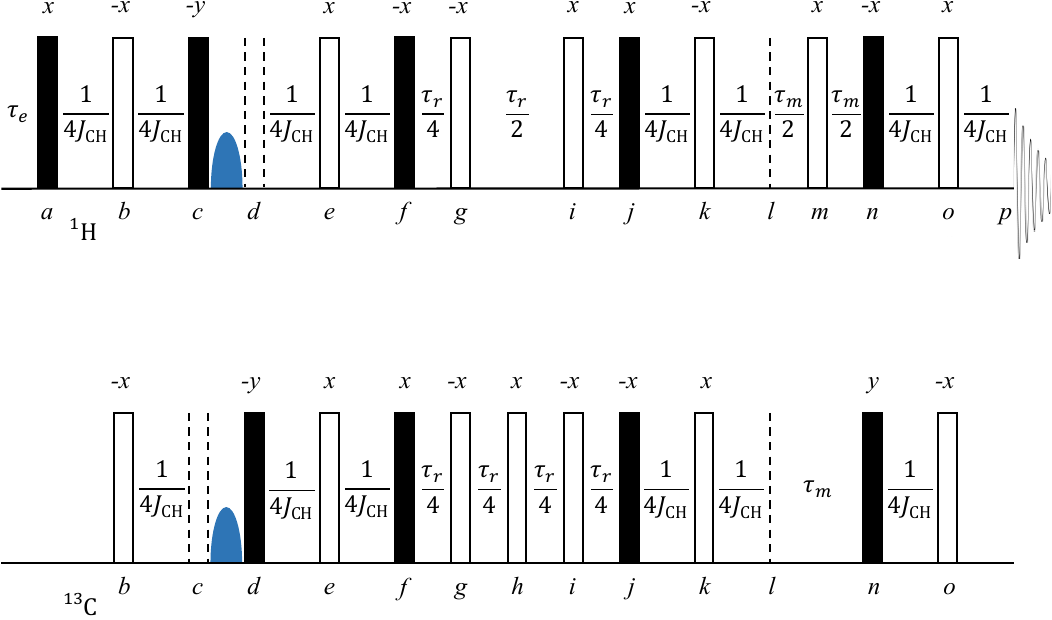}}
\caption{Proton-Proton Multiple Quantum Relaxation Pulse Sequence}
\label{fig:ProtonProtonMultipleQuantumPulseSequence}
\end{figure}

The pulse sequence depicted in Figure \ref{fig:ProtonProtonMultipleQuantumPulseSequence} observes both proton-proton multiple quantum relaxation and carbon longitudinal relaxation for ${}^{13}{\textrm{CH}}_2$. For ${}^{13}\textrm{CH}$, it observes only carbon longitudinal relaxation. For ${}^{13}{\textrm{CH}}_3$, it observes only proton-proton multiple quantum relaxation. It inserts proton-proton multiple quantum relaxation and/or carbon longitudinal relaxation into an HSQC experiment.

 \newpage
 
 The pulse sequence perturbs ${}^{13}{\textrm{CH}}_2$ as follows. The pulse sequence begins with a delay time $\tau_e$ that allows the spin system to achieve the thermal equilibrium state $4\hat{I_z}+\hat{S_z}$. A dephasing pulse sequence applied during time interval $a$-$c$ creates $2\hat{I_x}\hat{I_e}\hat{S_z} + 2\hat{I_e}\hat{I_x}\hat{S_z}$ anti-phase proton SQCs. A proton $90_{-y}^{\circ}$ pulse applied at time $c$ creates a $-2\hat{I_z}\hat{I_e}\hat{S_z} - 2\hat{I_e}\hat{I_z}\hat{S_z}$ $J$-ordered state. A purge gradient applied during time interval $c$-$d$ suppresses the $4\hat{I_y}$ SQCs for all protons that are not coupled to ${}^{13}\textrm{C}$. At time $d$, a carbon $90_{-y}^{\circ}$ pulse creates $-2\hat{I_z}\hat{I_e}\hat{S_x} - 2\hat{I_e}\hat{I_z}\hat{S_x}$ anti-phase polarization-enhanced carbon SQCs and $-\hat{S_x}$ in-phase non-enhanced carbon SQCs.
 
 A rephasing/dephasing pulse sequence applied during time interval $d$-$f$ rephases $-2\hat{I_z}\hat{I_e}\hat{S_x} - 2\hat{I_e}\hat{I_z}\hat{S_x}$ to $4\hat{S_y}$ polarization-enhanced in-phase SQCs and dephases $-\hat{S_x}$ to $\hat{I_z}\hat{I_z}\hat{S_y}$ doubly anti-phase SQCs and $-\hat{S_x} / 2$ in-phase SQCs. At time $f$, proton $90_{-x}^{\circ}$ and carbon $90_{x}^{\circ}$ pulses leave $-\hat{S_x} / 2$ unchanged and create $-4\hat{S_z}$ polarization-enhanced carbon $z$-magnetization, $-\hat{I_y}\hat{I_e}\hat{S_z} - \hat{I_e}\hat{I_y}\hat{S_z}$ anti-phase proton SQCs, $\hat{I_y}\hat{I_y}\hat{S_x}$ triple-quantum coherences (TQCs), and $-\hat{I_y}\hat{I_y}\hat{S_z}$ proton-proton MQCs, i.e., zero-quantum coherences (ZQCs) and double-quantum coherences (DQCs).
 
During time interval $f$-$j$, the spin system relaxes for a variable delay time $\tau_{r}$. Proton and carbon $180_{\pm{x}}^{\circ}$ pulses applied during this time interval refocus at time $j$ the effects of the proton and carbon chemical shifts and the $J_{\mathrm{CH}}$ scalar coupling. At time $j$, proton $90_{x}^{\circ}$ and carbon $90_{-x}^{\circ}$ pulses reconvert the SQCs, MQCs, and TQCs to $-4\hat{S_y}$ and $-\hat{I_z}\hat{I_z}\hat{S_y}$. Also, the proton $90_{x}^{\circ}$ converts $4\hat{I_z}$, which developed due to relaxation, to $4\hat{I_y}$ that becomes unobservable MQCs and TQCs at time $p$.
 
 A dephasing pulse sequence applied during time interval $j$-$l$ reconverts $-4\hat{S_y}$ and $-\hat{I_z}\hat{I_z}\hat{S_y}$ to $-2\hat{I_z}\hat{I_e}\hat{S_x} - 2\hat{I_e}\hat{I_z}\hat{S_x}$ and $-\hat{S_x}$ at time $l$, i.e., the same state as at time $d$. The remainder of the proton-proton multiple quantum relaxation pulse sequence applied during time interval $l$-$p$ achieves a similar result to time interval $i$-$o$ of the carbon-proton multiple quantum relaxation pulse sequence depicted in Figure \ref{fig:CarbonProtonMultipleQuantumPulseSequence}.

For relaxation delays $\tau_r=0$ and $\tau_r=\infty$, the $\alpha$ and $\beta$ components of the proton transverse magnetization on the $x$-axis (i.e., the proton doublet) at time $o$ are 

\begin{equation}
\begin{matrix}
 \alpha_{\tau_r={ \scriptscriptstyle 0} } = 2\cos \left ( \omega_{\mathrm{C}} \tau_{m} \right ) & \alpha_{\tau_r=\infty}= \frac {1} {4} \cos \left ( \omega_{\mathrm{C}} \tau_{m} \right )
 \\ 
 \beta_{\tau_r={ \scriptscriptstyle 0} } = 2\cos \left ( \omega_{\mathrm{C}} \tau_{m} \right ) & \beta_{\tau_r=\infty}= \frac {1} {4} \cos \left ( \omega_{\mathrm{C}} \tau_{m} \right )\end{matrix}
\label{eq:Cos5}
\end{equation}

For the AX spin system and relaxation delays $\tau_r=0$ and $\tau_r=\infty$, the $\alpha$ and $\beta$ components of the proton transverse magnetization on the $x$-axis at time $o$ are

\begin{equation}
\begin{matrix}
 \alpha_{\tau_r={ \scriptscriptstyle 0} } = 2\cos \left ( \omega_{\mathrm{C}} \tau_{m} \right ) & \alpha_{\tau_r=\infty}= \frac {1} {2} \cos \left ( \omega_{\mathrm{C}} \tau_{m} \right )
 \\ 
 \beta_{\tau_r={ \scriptscriptstyle 0} } = 2\cos \left ( \omega_{\mathrm{C}} \tau_{m} \right ) & \beta_{\tau_r=\infty}= \frac {1} {2} \cos \left ( \omega_{\mathrm{C}} \tau_{m} \right )\end{matrix}
\label{eq:Cos6}
\end{equation}

For the AX$_3$ spin system and relaxation delays $\tau_r=0$ and $\tau_r=\infty$, the $\alpha$ and $\beta$ components of the proton transverse magnetization on the $x$-axis at time $o$ are

\begin{equation}
\begin{matrix}
 \alpha_{\tau_r={ \scriptscriptstyle 0} } = 2\cos \left ( \omega_{\mathrm{C}} \tau_{m} \right ) & \alpha_{\tau_r=\infty}= 0
 \\ 
 \beta_{\tau_r={ \scriptscriptstyle 0} } = 2\cos \left ( \omega_{\mathrm{C}} \tau_{m} \right ) & \beta_{\tau_r=\infty}= 0
\end{matrix}
\label{eq:Cos7}
\end{equation}

\section*{Discussion}
\label{sec:Discussion}

The combination of the five coupled relaxation pulse sequences proposed above perturbs all 16 elements of the 4-by-4 density matrix for the AX spin system away from thermal equilibrium. The perturbation and subsequent temporal evolution of the density matrix have been validated by density matrix simulations of the pulse sequences, which were performed using the Maple programming language \cite{Maple}.

${}^{13}\textrm{C}$-coupled relaxation studies have revealed that perturbing a greater number of density matrix elements decreases the variances of fitting parameter values obtained via least-squares fits of experimental spectra to simulated spectra \cite{Liu}. Hence, the five pulse sequences that together perturb all 16 density matrix elements are expected to minimize the variances of fitting parameter values.

Another way to estimate the effect of each pulse sequence on the variance of fitting parameter values is to identify the spectral density functions that are used in the equation for each Redfield matrix element. Table \ref{tab:spectral_densities} shows the spectral density functions that contribute to Redfield relaxation of density matrix elements during the delay time $\tau_r$ for the AX spin system and for the carbon-proton multiple quantum, proton longitudinal, proton transverse, and carbon transverse relaxation pulse sequences.

\begin{table}[h!]
\renewcommand{\arraystretch}{1.35}
\begin{tabular}{lllll}
spectral density & carbon-proton MQ & proton long. & proton tran. & carbon tran. \\
$J_\mathrm{C,C}^\mathrm{CC} \left ( 0 \right) $ & \;\;\;\;\;\;\;\;\checkmark &  &  & \;\;\;\;\;\;\;\;\checkmark \\
$K_\mathrm{C,H}^\mathrm{CC} \left ( 0 \right) $ & \;\;\;\;\;\;\;\;\checkmark &   &  &   \\
$J_\mathrm{H,H}^\mathrm{CC} \left ( 0 \right) $ & \;\;\;\;\;\;\;\;\checkmark &  & \;\;\;\;\;\;\;\;\checkmark &  \\
$J_\mathrm{C,C}^\mathrm{CC} \left ( \omega_\mathrm{C} \right) $ & \;\;\;\;\;\;\;\;\checkmark & \;\;\;\;\;\;\;\;\checkmark & \;\;\;\;\;\;\;\;\checkmark & \;\;\;\;\;\;\;\;\checkmark \\
$J_\mathrm{C,C}^\mathrm{CC} \left ( \omega_\mathrm{H} \right) $ & \;\;\;\;\;\;\;\;\checkmark & \;\;\;\;\;\;\;\;\checkmark & \;\;\;\;\;\;\;\;\checkmark &\;\;\;\;\;\;\;\;\checkmark \\
$J_\mathrm{CH,CH}^\mathrm{DD} \left ( 0 \right) $ &  &  & \;\;\;\;\;\;\;\;\checkmark & \;\;\;\;\;\;\;\;\checkmark \\
$J_\mathrm{CH,CH}^\mathrm{DD} \left ( \omega_\mathrm{H} + \omega_\mathrm{C} \right) $ & \;\;\;\;\;\;\;\;\checkmark & \;\;\;\;\;\;\;\;\checkmark & \;\;\;\;\;\;\;\;\checkmark & \;\;\;\;\;\;\;\;\checkmark \\
$J_\mathrm{CH,CH}^\mathrm{DD} \left ( \omega_\mathrm{H} - \omega_\mathrm{C} \right) $ & \;\;\;\;\;\;\;\;\checkmark & \;\;\;\;\;\;\;\;\checkmark & \;\;\;\;\;\;\;\;\checkmark & \;\;\;\;\;\;\;\;\checkmark \\
$J_\mathrm{CH,CH}^\mathrm{DD} \left ( \omega_\mathrm{C} \right) $ & \;\;\;\;\;\;\;\;\checkmark & \;\;\;\;\;\;\;\;\checkmark & \;\;\;\;\;\;\;\;\checkmark & \;\;\;\;\;\;\;\;\checkmark \\
$J_\mathrm{CH,CH}^\mathrm{DD} \left ( \omega_\mathrm{H} \right) $ & \;\;\;\;\;\;\;\;\checkmark & \;\;\;\;\;\;\;\;\checkmark & \;\;\;\;\;\;\;\;\checkmark & \;\;\;\;\;\;\;\;\checkmark \\
$K_\mathrm{CH,C}^\mathrm{DC} \left ( 0 \right) $ & \;\;\;\;\;\;\;\;\checkmark &  & \;\;\;\;\;\;\;\;\checkmark & \;\;\;\;\;\;\;\;\checkmark \\
$K_\mathrm{CH,H}^\mathrm{DC} \left ( 0 \right) $ & \;\;\;\;\;\;\;\;\checkmark &  & \;\;\;\;\;\;\;\;\checkmark & \;\;\;\;\;\;\;\;\checkmark \\
$K_\mathrm{CH,C}^\mathrm{DC} \left ( \omega_\mathrm{C} \right) $ &  & \;\;\;\;\;\;\;\;\checkmark &  & \;\;\;\;\;\;\;\;\checkmark \\
$K_\mathrm{CH,H}^\mathrm{DC} \left ( \omega_\mathrm{H} \right) $ &  & \;\;\;\;\;\;\;\;\checkmark & \;\;\;\;\;\;\;\;\checkmark &
\end{tabular}
\caption{\label{tab:spectral_densities}Spectral density functions that contribute to relaxation during $\tau_r$}
\end{table}

The symbols in Table \ref{tab:spectral_densities} are interpreted respectively as follows. $J$ and $K$ specify auto-correlation and cross-correlation spectral density functions; the superscripts C and D specify chemical-shift-anisotropy and dipole-dipole relaxation; and the subscripts CH, C, and H specify a carbon-hydrogen dipole, carbon, and hydrogen.

Table \ref{tab:spectral_densities} reveals that, for the carbon-proton multiple quantum relaxation pulse sequence, all of the spectral density functions contribute to relaxation during the delay time $\tau_r$ except for $J_\mathrm{CH,CH}^\mathrm{DD} \left ( 0 \right) $, $K_\mathrm{CH,C}^\mathrm{DC} \left ( \omega_\mathrm{C} \right) $, and $K_\mathrm{CH,H}^\mathrm{DC} \left ( \omega_\mathrm{H} \right) $, which contribute to relaxation for the proton transverse and/or carbon transverse relaxation pulse sequences. However, Figure \ref{fig:CarbonProtonMultipleQuantumPulseSequence} reveals that proton transverse relaxation and carbon transverse relaxation occur during time intervals $a$-$c$ and $i$-$m$ respectively. Hence, for the carbon-proton multiple quantum relaxation pulse sequence, all of the spectral density functions contribute to relaxation, so this pulse sequence alone may suffice to obtain acceptable variances for fitting parameter values. 

All five relaxation pulse sequences are identical for the AX, AX$_2$, and AX$_3$ spin systems except that, for the carbon transverse and proton-proton relaxation pulse sequences, the optimum dephasing and rephasing delays for the AX$_2$ spin system during time intervals $d$-$f$, $i$-$k$, and $j$-$l$ are $ \frac{1} {8J_\mathrm{CH}} $ instead of $ \frac{1} {4J_\mathrm{CH}} $. If $ \frac{1} {4J_\mathrm{CH}} $ delays are used for the carbon transverse relaxation pulse sequence, $ \alpha_{\tau_r={ \scriptscriptstyle 0} } = \beta_{\tau_r={ \scriptscriptstyle 0} } = \cos \left ( \omega_{\mathrm{C}} \tau_{m} \right ) $ instead of $ \frac {3} {2} \cos \left ( \omega_{\mathrm{C}} \tau_{m} \right ) $ at time $o$ due to suppression of TQCs and carbon-proton DQCs by the purge gradient applied during time interval $h$-$i$. If $ \frac{1} {4J_\mathrm{CH}} $ delays are used for the proton-proton multiple quantum relaxation pulse sequence, carbon-proton ZQCs and DQCs are created instead of proton-proton ZQCs and DQCs at time $f$.

\section*{Conclusion}
\label{sec:Conclusion}

\emph{Spinach} programs that implement the pulse sequences depicted in Figures \ref{fig:CarbonProtonMultipleQuantumPulseSequence}-\ref{fig:ProtonTransversePulseSequence} are provided to facilitate interpretation of NMR coupled relaxation experiments.  \emph{Spinach} is a \emph{MatLab} library, so the nonlinear least squares curve fitting capability of \emph{MatLab} \cite{MatLab} could be used to fit experimental ${}^{13}\textrm{C}$-coupled relaxation spectra to simulated spectra generated by \emph{Spinach} in order to estimate molecular motion parameters such as rotational diffusion coefficients and rates of conformational change.

\section*{Supplemental Materials}

Ancillary files provided with this manuscript include source code for Maple programs that (1) perform density matrix simulations of the pulse sequences for the AX, AX$_2$, and AX$_3$ spin systems, (2) diagonalize the Hamiltonian matrices for these spin systems, and (3) derive equations for the Redfield matrix elements in terms of spectral density functions for these spin systems. Also provided are \emph{Spinach} source code that implements the pulse sequences depicted in Figures \ref{fig:CarbonProtonMultipleQuantumPulseSequence}-\ref{fig:ProtonTransversePulseSequence} and C++ source code that generates simulated proton spectra via Redfield relaxation of the AX spin system.

\newpage

\section*{Author Contact Information}

\href{https://www.linkedin.com/in/russellabrown/}{https://www.linkedin.com/in/russellabrown/}

\section*{Appendix}

Simulated 500 MHz proton spectra have been generated via Redfield relaxation of an isolated ${}^{13}\textrm{CH}$ spin system for the carbon-proton multiple quantum relaxation pulse sequence, proton longitudinal relaxation pulse sequence, and proton transverse relaxation pulse sequence depicted in Figures \ref{fig:CarbonProtonMultipleQuantumPulseSequence}-\ref{fig:ProtonTransversePulseSequence} respectively. The spectra were generated for a symmetric top rotating at $10^{10}$ radians per second with  $J_\mathrm{CH}$ =142 Hz, carbon CSA = 252 ppm, and proton CSA = 1 ppm.

\begin{figure}[h]
\centering
\centerline{\includegraphics[width=5.4in]{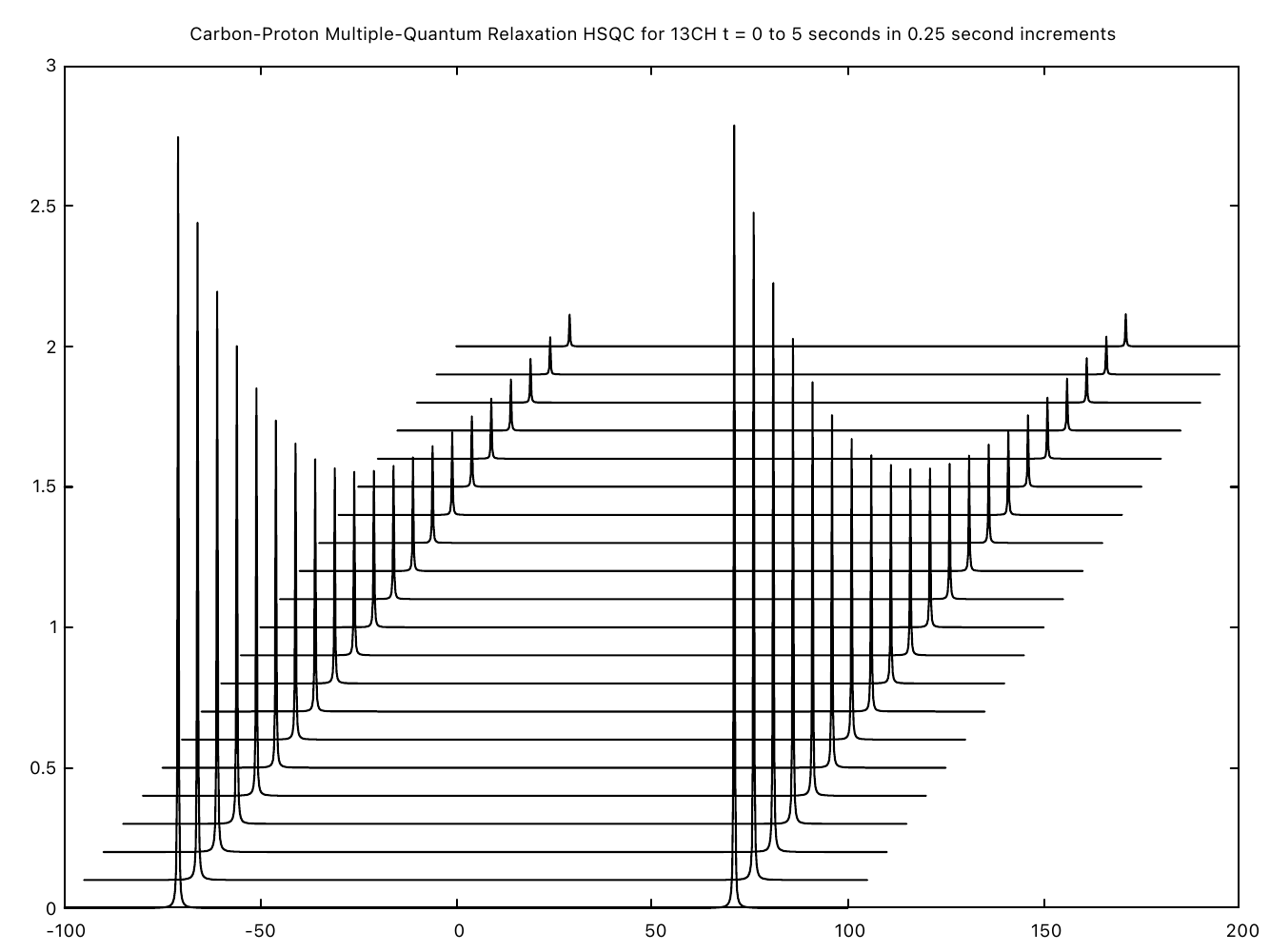}}
\caption{Carbon-Proton Multiple Quantum Relaxation Spectra}
\label{fig:CarbonProtonMultipleQuantumSpectrum}
\end{figure}

Figure \ref{fig:CarbonProtonMultipleQuantumSpectrum} plots the proton spectra generated for the carbon-proton multiple quantum relaxation pulse sequence depicted in Figure \ref{fig:CarbonProtonMultipleQuantumPulseSequence}.

\newpage

\begin{figure}[h]
\centering
\centerline{\includegraphics[width=5.4in]{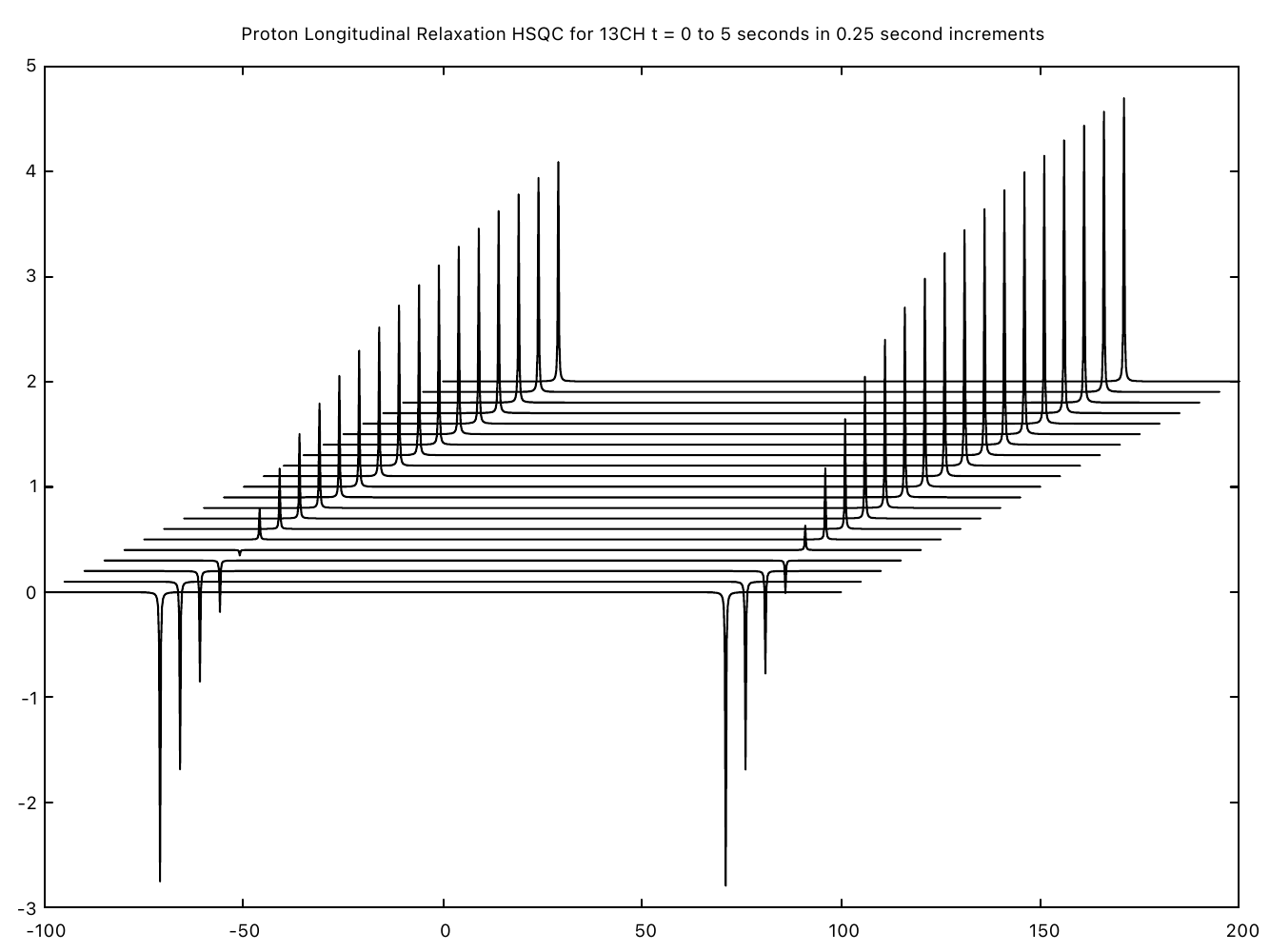}}
\caption{Proton Longitudinal Relaxation Spectra}
\label{fig:ProtonLongitudinalSpectrum}
\end{figure}

Figure \ref{fig:ProtonLongitudinalSpectrum} plots the proton spectra generated for the proton longitudinal relaxation pulse sequence depicted in Figure \ref{fig:ProtonLongitudinalPulseSequence}.

\newpage

\begin{figure}[h]
\centering
\centerline{\includegraphics[width=5.4in]{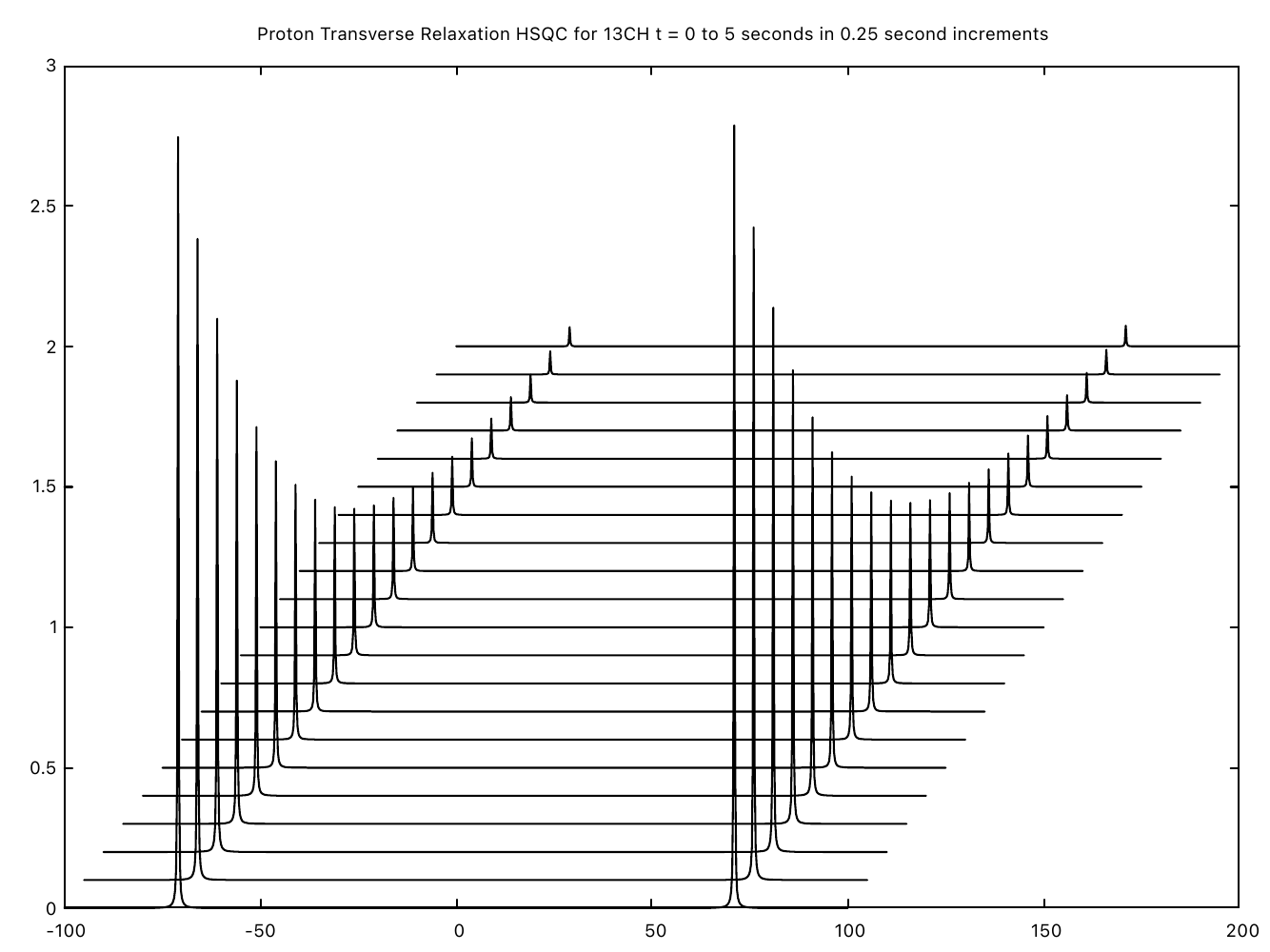}}
\caption{Proton Transverse Relaxation Spectra}
\label{fig:ProtonTransverseSpectrum}
\end{figure}

Figure \ref{fig:ProtonTransverseSpectrum} plots the proton spectra generated for the proton transverse relaxation pulse sequence depicted in Figure \ref{fig:ProtonTransversePulseSequence}. Although these spectra resemble the spectra for the carbon-proton multiple quantum pulse sequence depicted in Figure \ref{fig:CarbonProtonMultipleQuantumPulseSequence}, comparison of their spectra reveals that the relaxation rates differ for these two pulse sequences.

\newpage

\section*{References}

\small
\bibliographystyle{jcgt}
\bibliography{paper}

@ARTICLE{Werbelow,
   AUTHOR  = {L.G. Werbelow and D.M. Grant},
   TITLE   = {Carbon-13 relaxation in multispin systems of the type {AX}$_{n}$},  
   YEAR    = {1975},
   JOURNAL = {Journal of Chemical Physics},
   VOLUME  = {63},
   NUMBER  = {1},
   PAGES   = {544-556},
   URL = {https://pubs.aip.org/aip/jcp/article-abstract/63/1/544/784016/Carbon-13-relaxation-in-multispin-systems-of-the?redirectedFrom=fulltext}
}

@ARTICLE{Favro,
   AUTHOR  = {L.D. Favro},
   TITLE   = {Theory of the rotational {B}rownian motion of a free rigid body},  
   YEAR    = {1960},
   JOURNAL = {Physical Review},
   VOLUME  = {119},
   NUMBER  = {1},
   PAGES   = {53-62},
   URL = {https://journals.aps.org/pr/abstract/10.1103/PhysRev.119.53}
}

@ARTICLE{Ryabov,
   AUTHOR  = {Y. Ryabov and G.M. Clore and C.D. Schwieters},
   TITLE   = {Coupling between internal dynamics and rotational diffusion in the presence of exchange between discrete molecular conformations},  
   YEAR    = {2012},
   JOURNAL = {Journal of Chemical Physics},
   VOLUME  = {136},
   NUMBER  = {},
   PAGES   = {034108},
   URL = {https://pubs.aip.org/aip/jcp/article-abstract/136/3/034108/190943/Coupling-between-internal-dynamics-and-rotational?redirectedFrom=fulltext}
}

@INCOLLECTION{Huntress,
   AUTHOR = {W.T. Huntress},
   TITLE = {The study of anisotropic rotation of molecules in liquids by {NMR} quadrupolar relaxation},
   BOOKTITLE = {Advances in Magnetic and Optical Resonance}, 
   EDITOR = {J.S. Waugh},
   VOLUME  = {4},
   PAGES = {1-37}, 
   YEAR = 1970, 
   publisher = {Academic Press},
   address = {New York, NY},
   URL = {https://www.sciencedirect.com/science/article/abs/pii/B9780120255047500076}
}

@ARTICLE{Redfield,
   AUTHOR  = {A.G. Redfield},
   TITLE   = {The theory of relaxation processes},  
   YEAR    = {1965},
   JOURNAL = {Advances in Magnetic and Optical Resonance},
   VOLUME  = {1},
   NUMBER  = {},
   PAGES   = {1-32},
   URL = {https://www.sciencedirect.com/science/article/abs/pii/B9781483231143500076}
}

@ARTICLE{Fuson,
   AUTHOR  = {M.M. Fuson and A.M. Belu},
   TITLE   = {Coupled-spin relaxation of {AX}$_{2}$ spin systems in the presence of neighboring spins},  
   YEAR    = {1994},
   JOURNAL = {Journal of Magnetic Resonance},
   VOLUME  = {A107},
   NUMBER  = {1},
   PAGES   = {1-7},
   URL = {https://www.sciencedirect.com/science/article/abs/pii/S1064185884710400}
}

@ARTICLE{Brown1995,
   AUTHOR  = {R.A. Brown and D.M. Grant},
   TITLE   = {${}^{13}\textrm{C}$-coupled relaxation studies of a leucine zipper peptide using polarization-transfer pulse sequences},  
   YEAR    = {1995},
   JOURNAL = {Journal of Magnetic Resonance},
   VOLUME  = {B106},
   NUMBER  = {3},
   PAGES   = {253-260},
   URL = {https://www.sciencedirect.com/science/article/abs/pii/S1064186685710412}
}

@ARTICLE{Liu,
   AUTHOR  = {F. Liu and C.L. Mayne and D.M. Grant},
   TITLE   = {Magnetization preparation for coupled relaxation studies using ${J}$-spectral pulse sequences},  
   YEAR    = {1989},
   JOURNAL = {Journal of Magnetic Resonance (1969)},
   VOLUME  = {84},
   NUMBER  = {2},
   PAGES   = {344-350},
   URL = {https://www.sciencedirect.com/science/article/abs/pii/0022236489903776}
}

@ARTICLE{Zheng,
   AUTHOR  = {Z. Zheng and C.L Mayne and D.M. Grant},
   TITLE   = {Ethanol molecular dynamics measured by coupled spin relaxation exhibiting cross correlation between dipole-dipole and chemical-shift anisotropy},  
   YEAR    = {1993},
   JOURNAL = {Journal of Magnetic Resonance},
   VOLUME  = {A103},
   NUMBER  = {3},
   PAGES   = {268-281},
   URL = {https://www.sciencedirect.com/science/article/abs/pii/S1064185883711666}
}

@ARTICLE{Kuprov2011,
   AUTHOR  = {I. Kuprov},
   TITLE   = {Diagonalization-free implementation of spin relaxation theory for large spin systems},  
   YEAR    = {2011},
   JOURNAL = {Journal of Magnetic Resonance},
   VOLUME  = {209},
   NUMBER  = {1},
   PAGES   = {31-38},
   URL = {https://www.sciencedirect.com/science/article/abs/pii/S1090780710003964}
}

@ARTICLE{Kuprov2021,
   AUTHOR  = {I. Kuprov and L.C. Morris and J.N. Glushka and J.H. Prestegard},
   TITLE   = {Using molecular dynamics trajectories to predict nuclear spin relaxation behaviour in large spin systems},  
   YEAR    = {2021},
   JOURNAL = {Journal of Magnetic Resonance},
   VOLUME  = {323},
   PAGES   = {106891},
   URL = {https://www.sciencedirect.com/science/article/abs/pii/S1090780720302093}
}

@MISC{Spinach,
   AUTHOR = {Spinach},
   BOOKTITLE = {SPINACH Library for MATLAB},
   URL = {https://spindynamics.org/},
   YEAR = {2023},
   }

@MISC{MatLab,
   AUTHOR = {MathWorks},
   BOOKTITLE = {Nonlinear Least Squares Curve Fitting},
   URL = {https://www.mathworks.com/help/optim/nonlinear-least-squares-curve-fitting.html},
   YEAR = {2023},
   }

@MISC{Maple,
   AUTHOR = {Maplesoft},
   BOOKTITLE = {Maple Symbolic Algebra Software},
   URL = {https://www.maplesoft.com/},
   YEAR = {2023},
   }

@INCOLLECTION{Keeler1,
   AUTHOR = {J. Keeler},
   TITLE = {Selection of $z$-magnetization},
   BOOKTITLE = {Understanding NMR Spectroscopy}, 
   EDITION = {Second},
   PAGES = {424-425}, 
   YEAR = {2010}, 
   PUBLISHER = {John Wiley and Sons, Ltd},
}

@INCOLLECTION{Keeler2,
   AUTHOR = {J. Keeler},
   TITLE = {{HSQC}},
   BOOKTITLE = {Understanding NMR Spectroscopy}, 
   EDITION = {Second},
   PAGES = {420}, 
   YEAR = {2010}, 
   PUBLISHER = {John Wiley and Sons, Ltd},
}

@INCOLLECTION{Keeler3,
   AUTHOR = {J. Keeler},
   TITLE = {Sensitivity-enhanced {HSQC}},
   BOOKTITLE = {Understanding NMR Spectroscopy}, 
   EDITION = {Second},
   PAGES = {350-353}, 
   YEAR = {2010}, 
   PUBLISHER = {John Wiley and Sons, Ltd},
}

@INCOLLECTION{Keeler4,
   AUTHOR = {J. Keeler},
   TITLE = {{TROSY}},
   BOOKTITLE = {Understanding NMR Spectroscopy}, 
   EDITION = {Second},
   PAGES = {358-366}, 
   YEAR = {2010}, 
   PUBLISHER = {John Wiley and Sons, Ltd},
}

@ARTICLE{Pervushin,
   AUTHOR  = {K. Pervushin and R. Riek and G. Wider and K. Wutrich},
   TITLE   = {Attenuated ${T_{2}}$ relaxation by mutual cancellation of dipole-dipole coupling and chemical shift anisotropy indicates an avenue to {NMR} structures of very large biological macromolecules in solution},  
   YEAR    = {1997},
   JOURNAL = {Proceedings of the National Academy of Science, USA},
   VOLUME  = {94},
   NUMBER  = {23},
   PAGES   = {12366-12371},
   URL = {https://www.pnas.org/doi/10.1073/pnas.94.23.12366}
}

@ARTICLE{Bax,
   AUTHOR  = {A. Bax and R.H. Griffey and B.L. Hawkins},
   TITLE   = {Correlation of proton and nitrogen-15 chemical shifts by multiple quantum NMR},  
   YEAR    = {1983},
   JOURNAL = {Journal of Magnetic Resonance (1969)},
   VOLUME  = {55},
   NUMBER  = {2},
   PAGES   = {301-315},
   URL = {https://www.sciencedirect.com/science/article/abs/pii/002223648390241X}
}

@ARTICLE{Bodenhausen,
   AUTHOR  = {G. Bodenhausen and D.J. Ruben},
   TITLE   = {Natural abundance nitrogen-15 {NMR} by enhanced heteronuclear spectroscopy},  
   YEAR    = {1980},
   JOURNAL = {Chemical Physics Letters},
   VOLUME  = {69},
   NUMBER  = {1},
   PAGES   = {185-189},
   URL = {https://www.sciencedirect.com/science/article/abs/pii/0009261480800418}
}

@ARTICLE{Morris,
   AUTHOR  = {G.A. Morris and R. Freeman},
   TITLE   = {Enhancement of nuclear magnetic resonance signals by polarization transfer},  
   YEAR    = {1979},
   JOURNAL = {Journal of the American Chemical Society},
   VOLUME  = {101},
   NUMBER  = {3},
   PAGES   = {760-762},
   URL = {https://pubs.acs.org/doi/pdf/10.1021/ja00497a058}
}

@ARTICLE{Burum,
   AUTHOR  = {D. Burum and R. Ernst},
   TITLE   = {Net polarization transfer via a ${J}$-ordered state for signal enhancement of low-sensitivity nuclei},  
   YEAR    = {1980},
   JOURNAL = {Journal Magnetic Resonance},
   VOLUME  = {39},
   NUMBER  = {1},
   PAGES   = {163-168},
   URL = {https://www.sciencedirect.com/science/article/abs/pii/0022236480901687}
}

\end{document}